# Pulsed DC bias for the study of negative-ion production on surfaces of insulating materials in low pressure hydrogen plasmas

K. Achkasov[1,2], R. Moussaoui[1], D. Kogut[1], E. Garabedian[1], J. M. Layet[1], A. Simonin[2], A. Gicquel[3], J. Achard[3], A. Boussadi[3] and G. Cartry[1]

[1]*Aix-Marseille Université, CNRS, PIIM, UMR 6633, Centre Scientifique de Saint Jérôme, case 241, 13397 Marseille Cedex 20, France*
[2]*CEA, IRFM, F-13108 Saint-Paul-lez-Durance, France*
[3]*Université Paris 13, Sorbonne Paris Cité, Laboratoire des Sciences des Procédés et des Matériaux, CNRS (UPR 3407), 93430 Villetaneuse, France*

## Abstract

In this work negative-ion production on the surface of a sample negatively DC biased in a hydrogen plasma is studied. The negative ions created under the positive ion bombardment are accelerated towards the plasma, self-extracted and detected according to their energy and mass, by a mass spectrometer placed in front of the sample. The use of a pulsed bias allows applying a quasi-DC bias on insulating material during a short period of time and offers the possibility to extend the measurement method to nonconductive samples. The pulsed-bias tests were performed first with Highly Oriented Pyrolitic Graphite (HOPG), a conductive material, to demonstrate the feasibility of the method. By changing the pulsed-bias frequency it was possible to obtain HOPG material with different hydrogen surface coverages and hence different surface states leading to an increase of negative-ion production by up to *30-50%* as compared to the continuous bias case. To establish a protocol for insulating materials, charge accumulation on the surface during the bias pulse and influence of the bias duration and frequency were explored using microcrystalline diamond (MCD) thin layers. By using a pulse short enough (10 µs) at 1 kHz frequency, it has been possible to measure negative-ions on MCD sample at a quasi-constant surface bias of 130 V, with only 1 V variation during the measurement. Negative-ion surface production on MCD has been studied in pulsed mode with surface temperature from room temperature to 800°C. It is shown that pulsing the bias and increasing the temperature allows limiting defect creation on MCD which is favorable for negative-ion production. Consequently, at 400°C the yield on MCD in pulsed mode is one order of magnitude higher than the yield on HOPG in continuous mode at room temperature.

Corresponding author: gilles.cartry@univ-amu.fr

## 1. Introduction

Negative ions (NI) in low-pressure plasmas are usually created by dissociative attachment of cold plasma electrons (< *1 eV*) on vibrationally excited molecules. This process is called NI volume production[1,2]. However, NI can also be produced by surface ionization of backscattered atoms at plasma reactor walls. In most experimental conditions, NI surface production is usually a mechanism of minor importance. However, in certain circumstances it can be very efficient. When alkali metal surfaces such as cesium (Cs) are in contact with plasma, a huge surface production of NI by conversion of positive ions or atoms is observed[3]. This effect is the basis of the most intense H$^-$/D$^-$ negative ion sources developed for fusion or particle accelerator applications[4,5,6,7]. However, the presence of cesium inside negative-ion sources complicates their operation and alternative solutions to produce high rate of negative ions on surfaces in contact with hydrogen plasma would be valuable. These alternatives include the use of low work-function materials other than cesium[8] or investigation of other materials with interesting electronic properties such as dielectric materials. Electron capture by incoming particles on dielectric material is harder due to the absence of electron in the conduction band. However, the transfer back of the captured electron from the negative-ion to the material is decreased or suppressed due to the presence of the gap[9]. This is resulting in an ionization efficiency that might be very high as demonstrated in beam experiments[10]. Despite the processes of electron capture and loss by incoming particles are different for dielectrics and low-work function materials due to different electronic structures[9], it is expected that the former could also give high negative ion yield when placed in contact with hydrogen plasma. Within this framework we investigate negative-ion surface production on carbon materials putting a strong focus on diamond which presents electronic properties favorable for surface ionization[9]. Negative-ion surface production on carbon materials may also be of interest in the context of plasma processes for the microelectronic industry where carbon materials may be in contact with electronegative gases.

In this study a sample is positioned inside the plasma facing a quadrupole mass spectrometer equipped with an energy filter. The sample is negatively biased with respect to the plasma potential. Positive ions bombard the sample and negative ions formed on the sample surface upon this bombardment are accelerated towards the plasma. They gain in the sheath in front of the sample enough energy to overcome the sheath potential in front of the mass spectrometer leading to their self-extraction from the plasma. Negative ions are detected according to their energy and mass by the mass spectrometer and Negative-Ion Energy Distribution Functions (NIEDF) are measured. The analysis of NIEDF shape gives information on negative-ion surface production[11,12]. To study NI surface production on conductive samples, a constant DC bias is applied. The DC bias leads to positive ion bombardment of the sample with a well-defined energy. It also permits the self-extraction of negative-ions created on the surface towards the mass spectrometer. When using the same technique for insulating samples, charge is quickly built up on the surface. As a result, the applied negative voltage does not appear anymore on the surface which is floating, so negative-ions can no more be self-extracted preventing the measurements of NIEDF. To eliminate the surface charging an RF bias is typically used but its capability to control the energy of the bombarding ions is limited[13] and furthermore it would strongly modify NIEDF shape and complicate their interpretation. The method of pulsed DC bias described in the present paper was developed to enable the study of NI production on surfaces of insulating materials using the same tools as those developed for conductive materials. The idea is to use a pulse which is short enough to get a small charge build-up during the pulse and thus a negligible voltage change, and use a period long enough after the pulse to let the plasma neutralize the charge on the surface. During the negative pulse, positive ions bombard the sample at well-defined energy, negative-ions are self-extracted and NIEDF measurements are performed. A similar method has been used by Samara et al[14] to measure the ion saturation current in a simpler manner as compared to the RF-biased probe. Wang and Wendt[15] and Kudlacek et al[13] have used this approach to control the ion energy in industrial plasma processes dealing with insulators. They have applied a modulated pulse shape, where the voltage during the pulse is sloped and thus exactly compensates for the drop of voltage due to the charging of the substrate being processed. In the present paper, only rectangular shape pulses were used such as in the study of Barnat et al[16] where pulse DC bias was employed for the sputtering of insulators.

Based on previous studies[17], an insulating microcrystalline diamond film exposed to hydrogen plasma has proven to be conducting only starting from 300°C. Below this temperature, the NI signal could not be measured as the DC bias could not be applied. The present technique enables to study NI surface production on diamond films for the whole temperature range starting from room temperature (RT) until 800°C. Moreover, this technique extends the measurement of NI surface production to other potentially interesting insulating materials.

The paper is separated in five parts. The first two concern the presentation of the materials used and the principle of the pulsed bias method. The third part is dedicated to results obtained in pulsed bias mode with a conductive sample while the fourth part is dedicated to an insulating diamond layer. The last part shows negative-ion signals obtained in pulsed mode versus surface temperature for several samples.

## 2. Materials

HOPG (Highly Oriented Pyrolitic Graphite) has been chosen as a reference material for the study because of its high negative-ion production rate[18,19,20] and because of its simplicity to cleave and clean. The HOPG material was of ZYB type purchased from MaTeck GmbH company. The density and electrical resistivity of HOPG were 2.265 g·cm$^{-3}$ and 3.5×10$^{-5}$ Ω·cm, respectively. Diamond is another perspective material for high NI surface production rate[9]. It is well known for its ability to emit electrons at high temperature and even at low electric fields[21]. Beam experiments on diamond showed surface production of H$^{-}$ ions with high yields up to *5.5%*[22] (yield is defined in these experiments as the negative ion fraction in the reflected particle flux). Moreover, it has been observed in plasma experiment that NI production rate on boron-doped-diamond can be increased by a factor of 5 when increasing the temperature to 400°C[23,24]. This has raised the interest to study NI production on heated surfaces in plasma.

Microcrystalline boron-doped-diamond (MCBDD) and microcrystalline non-doped-diamond (MCD) films used in this study were deposited at LSPM laboratory by using plasma-enhanced chemical vapor deposition (PECVD). The films were deposited on a <100> oriented 1 mm thick doped silicon substrate. The deposited diamond layer for MCBDD had a thickness of 3.2 ± 0.1 μm (determined by the weight gain of the substrate after deposition). For MCD the layer thickness was 17.5 ± 0.5 μm (determined from confocal microscopy pictures). The grain size for MCBDD samples is about 2.5 μm in the horizontal plane and much bigger for MCD, on the order of 10 μm, as seen by scanning electron and confocal microscopy pictures. The doping level for MCBDD was estimated to be 10$^{19}$ - 10$^{20}$ cm$^{-3}$, which leads to sufficiently good electrical conductivity for biasing of the diamond layer. Even if hydrogenated diamond surfaces are well known to be conductive, it has been observed that under plasma exposure at surface temperature lower than 300°C, MCD samples lose their surface conductivity[17], most probably because of the creation of a defective layer on top of the diamond thin film.

## 3. Principle of measurements and experimental set-up

An insulator sample biased by a DC pulse and immersed into plasma acts as a capacitor, which can be, in a first approach, considered as planar. The insulator sample surface is one plate of the capacitor with the surface potential $V_s$. The second face of the insulator is in contact with a polarizing plate, acting as the second plate of the capacitor with a potential equal to the applied one $V_a$. The surface potential is equal to:

$$V_s = Q/C + V_a = Q\,d\,/(\varepsilon_0 \cdot \varepsilon_r \cdot S) + V_a \qquad (1)$$

where $Q$ is the charge accumulated by the capacitor (in Coulomb) and $C$ is its capacitance. The capacitance can be calculated from the known vacuum permittivity *$\varepsilon_0$ = 8.85·10$^{-12}$ (F/m)*, relative permittivity of the insulating material $\varepsilon_r$, surface of the planar insulator $S$ and its thickness $d$.

In such a way, the surface bias time variation reads:

$$\frac{dV_s}{dt} = \frac{1}{C}\frac{dQ}{dt} \qquad (2)$$

When the pulse is applied, positive ions are attracted towards the surface. Here we consider that the applied bias $V_a$ is sufficiently negative so that the sample current is the positive ion saturation current. The ion saturation current $I_{isat}$ of the positive ions attracted towards the sample reads for a planar sheath:

$$I_{isat} = 0.6 \cdot e \cdot n_i \sqrt{\frac{kT_e}{m_i}} \cdot S \qquad (3)$$

where $e$ is electron charge, $n_i$ is plasma ion density, $m_i$ is ion mass, $T_e$ is electron temperature and $k$ is the Boltzmann constant. The sample current is constant as long as the surface bias $V_s$ remains much smaller than plasma potential $V_P$. Then, $V_s$ can be obtained by integrating its time variation, taking into account the initial condition of the surface bias before the pulse application ($V_{s\,(t=0)}$ is equal to the floating potential $V_f$). The surface bias $V_s$ varies linearly with time following equation (4).

$$V_s = \frac{I_{isat}}{C} \cdot t + V_a + V_f \qquad (4)$$

Figure 1 (a) sketches applied ($V_a$) and surface bias ($V_s$) with an insulating sample.

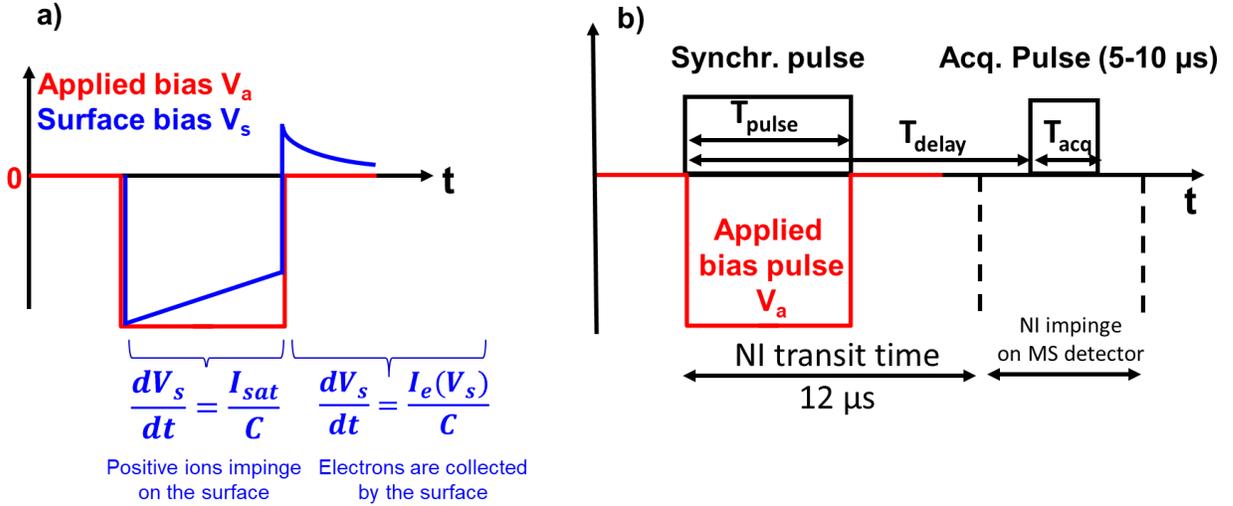

Figure 1. (a) Sketch of surface voltage behaviour during pulsed bias experiments (b) Sketch of the synchronization scheme.

When the pulse is over, the applied voltage is zero $V_a = 0$, so the surface bias becomes positive: $V_s = Q/C$. If the charge $Q$ is high enough, the surface bias becomes higher than the plasma potential $V_P$. In such a way, a huge electron flux is attracted towards the sample during the first instants due to high electron mobility. This causes a perturbation in plasma, resulting in a local increase of the plasma potential which limits the electron loss by setting the sample surface at a potential close to the (disturbed) floating potential. The sample is then discharged by electrons from the plasma on a longer time scale (as demonstrated in Figure 1a). If the initial charge Q is low and the surface bias is not higher than the plasma potential, there is no disturbance of the plasma and the sample discharge happens at potential close to the floating potential $V_f$.

So, when being negatively DC-biased, insulator surface accumulates positive ions on the surface. If one applies a periodic short-time negative DC-bias to the insulator, and no bias for the rest of the time, the positive charge accumulated during the bias period can be discharged by electrons during the unbiased period. The NI emitted from the sample can be measured during the pulse, so the measured signal can be compared directly to NIEDF obtained on conductive samples using continuous DC-bias. To set ideas on a simple example, let us consider a 5 µm thick 1 cm² diamond layer biased at -100 V in plasmas with ion fluxes in the range 1 to 1000 µA/cm². Under this situation, the voltage change rate during the pulse (equation 4) is in the range 1 mV/µs to 1 V/µs. As the time resolution of the mass spectrometer is 2 µs it allows for NIEDF acquisition with a voltage change of maximum 2% during the measurement. For higher positive ion flux (or higher sample thickness) a simple rectangular shape pulse can no more be used, and the technic proposed by Kudlacek *et al.* must be employed[13]. Since the voltage change rate during the pulse will be more important, in order to preserve a constant charge on the surface one would have to apply a modified pulse shape. The trapezoidal pulse shape with a linear voltage increase during the pulse should then exactly compensate for the drop of voltage due to the charging of the surface.

The measurements set-up is described elsewhere[25]. The plasma is created in the capacitive mode by an external antenna connected to a 13.56 MHz generator through a matchbox and surrounding a Pyrex tube. H$_2$ plasma is ignited under the following experimental conditions: 2 Pa, RF power of 26 W, surface bias *V$_s$ = – 130 V*, distance between the mass spectrometer and the sample holder *37 mm*. To minimize RF fluctuations of plasma potential arising due to capacitive coupling, a mechanical grounded screen was placed horizontally approximately 5 cm above the sample, between the source and the diffusion chamber[25]. Let us note that the injected power has been slightly increased compared to our previous works, from 20 to 26 W. This increase of the power was required to maintain the plasma density, as measured by Langmuir probe and inferred from sample current, at the level of our previous studies. Most probably a change of antenna to plasma coupling has occurred, maybe due to mechanical change in the position of the antenna. Anyway, plasma parameters are identical and results are directly comparable between the present work and our previous studies. The sample was held in place on the sample holder by a 2 mm thick molybdenum clamp or by a 3 mm thick ceramic clamp. The reason for using two different materials as clamp is to test the influence of the sheath shape in front of the sample on the results. The clamp material used for the measurements will be indicated in the figures. The radius of the clamp hole defining the sample area exposed to plasma was 4 mm. Under the considered experimental conditions the positive ion current density is on the order of ~ 10 µA/cm$^2$.

Samples are biased thanks to a DC voltage source connected to a homemade electronic chopper controlled by a function generator. The bias duration *T$_{pulse}$* and the bias frequency f are determined by the function generator. A delay generator synchronized with the function generator is used to define the delay T$_{delay}$ between the DC bias pulse and the acquisition, as well as the duration of the acquisition T$_{acq}$. The *12 µs* H$^-$ transit time between the sample and the mass spectrometer detector has been measured accurately by observing on a scope counts at the output of the mass spectrometer detector. *T$_{delay}$ = 13 µs* has been chosen to get rid of NI created during the first microsecond of the pulse when the applied bias is establishing. In the same way the acquisition duration *T$_{acq}$* was usually set shorter than the applied bias in order prevent the collection of NI created during the fall down of the sheath in front of the sample at the end of the pulse. This is summarized in Figure 1 (b).

In order to find optimal experimental conditions and to test the feasibility of the method at room temperature (RT), first tests were performed on HOPG as a reference material. Then, the optimization of the bias and acquisition parameters was performed on MCD by taking into account the charging effects. Finally, the results for the whole temperature range were obtained.

## 4. Results for a conductive sample: HOPG

Under the present experimental conditions, most of the negative ions detected have suffered no collision during their transport between the sample and the mass spectrometer detector. Therefore, the negative-ion total energy is conserved between the sample and the mass spectrometer and can be measured by the mass spectrometer energy filter:

$$E_T = E_{ks} - eV_s \qquad (5)$$

Where $E_T$ is the negative ion total energy and $E_{ks}$ is the kinetic energy at which the ion is created on the surface. Knowing the surface bias $V_s$, which is equal to the applied bias for conductive materials, NIEDF data can be plotted versus $E_{ks}$ as shown for instance in Figure 2 (a). In this representation zero energy corresponds to a negative ion created at rest on the surface. Any increase of the surface bias $V_s$ due to charging would lead to a shift of NIEDF towards negative values ($V_s$ is negative). In order to trace the possible variation of the surface bias $V_s$, the delay time $T_{delay}$ was varied from *13 µs* to *953 µs* using a 1 ms pulse ($T_{pulse}$ *= 1 ms*) at 100 Hz (duty cycle 10%) and an acquisition duration of 50 µs ($T_{acq}$ *= 50 µs*) (not shown here). The peak position of NIEDFs was unchanged for all the measurements, which means that no charging effects are present with HOPG material.

The change of the bias frequency has proven to influence the hydrogen surface dynamics. The total NI signal, defined as the area below NIEDF, is plotted versus the pulse frequency in arbitrary units in Figure 2 (b). The value of NI signal for constant bias and constant acquisition has been taken as a reference and normalized to 1. The other data have been normalized accordingly. It can be seen that NI signal in pulsed mode is approximately twice the reference level at low frequency and decreases with the increase of frequency. The full symbols represent the results for constant DC-bias, but pulsed mass spectrometer acquisition. As one can see, the NI yield stays nearly constant for different acquisition frequencies as expected. The slight decrease observed is attributed to experimental uncertainties rather to a physical effect. Therefore, it can be concluded that the high NI signal at low pulse frequency and its decrease with frequency during the pulsed bias experiment is connected to the surface state change of the material and not to acquisition issues. The important point of the graph is that pulsed bias results in enhanced NI production as compared to constant DC-bias.

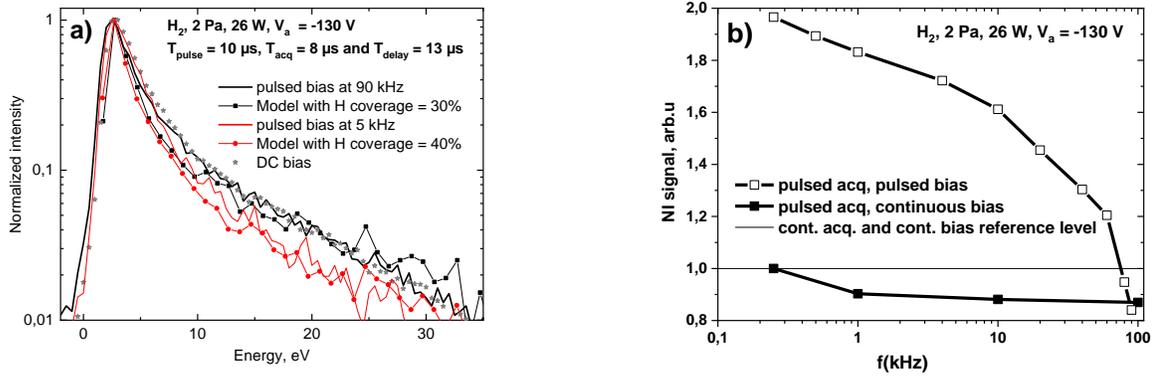

Figure 2: H$_2$, 2.0 Pa, 26 W, Va = Vs = -130 V, Tpulse = 10 μs, Tacq = 8 μs and Tdelay = 13 μs, molybdenum clamp (a) Normalized NIEDF measured on HOPG at RT for constant DC-bias (star symbols) and for pulsed-bias at two different frequencies (black and red lines). Normalized NIEDF computed for two different H atom surface coverage (30% black square symbols, 40% red circle symbols) (b) NI total signal dependence for HOPG with the pulsed bias frequency. The open square symbols represent the result for pulsed DC bias and pulsed mass spectrometer acquisition. The closed square symbols represent the result for continuous DC bias and pulsed mass spectrometer acquisition. The DC bias signal has been taken as a reference level and set to 1. Pulsed acquisitions are normalized accordingly.

Figure 2 (a) reveals that the normalized Negative Ion Energy Distribution Functions (NIEDF) change in shape with pulse frequency. The NIEDF energetic tail is growing with increasing pulse frequency. It was verified that for low frequencies (from *0.25* to *1 kHz*) the shape of the NIEDF was approximately the same. By analyzing NIEDF, one could get a better idea of what is happening with HOPG surface under different pulsed bias conditions

In previous works it was demonstrated that negative-ions are created by backscattering of positive-ions after their neutralization and the capture of an electron on the surface, and by sputtering of adsorbed hydrogen as a negative-ion[18,19,20]. A model based on the SRIM software has been developed to determine negative-ion angle and energy distribution functions on the surface. SRIM is a software package which calculates many features related to ion impact on surfaces such as implantation, backscattering or sputtering[26]. In particular, SRIM is able to output angle and energy distributions of sputtered and backscattered particles upon impact of a given ion on a certain surface. Input parameters are ion type and energy, as well as surface state definition. In the present case a hydrogenated carbon surface a-C(H) has been defined with parameters obtained from the literature[28].

It has been assumed that the backscattered and sputtered distribution functions obtained from SRIM are valid for negative-ions. Based on SRIM outputs (energy and angle of emitted particles), negative-ions trajectories in the sheaths and in the plasma are computed using

Newton's law of motion. Ions missing the mass spectrometer or arriving at angles larger than the acceptance angle are eliminated from the calculation. Then, thanks to the SIMION[27] software, negative ion transmission function through the mass spectrometer is estimated to finally get calculated NIEDF at mass spectrometer detector[11,12, 25]. Calculated NIEDF are compared to experimental ones. The hydrogen surface coverage of the a-C(H) layer has a strong influence on calculated NIEDF. It has been estimated using previous publications and has then been varied until the model matches the experiment[28]. Once the calculated NI distribution matches the experimental one, it is assumed that the distribution of particles given by SRIM is those of negative ions.

The model has been benchmarked versus the experiments and it has been shown[9,25] that an increase of the NIEDF tail with respect to the peak of the NIEDF corresponds to an increase of the backscattering contribution to the total NI production and to a decrease of the sputtering contribution. Indeed, NI created through sputtering of adsorbed H atoms are mostly produced at low energy while positive ions backscattered as NI (after neutralization and electron capture) are distributed over the whole energy range, from 0 to the maximum energy determined by the sheath drop voltage: $e(V_p-V_s)$ ($V_p$ is the plasma potential)[28].

A decrease of the sputtering contribution at constant positive ion energy corresponds to a decrease of the hydrogen top surface coverage. To support this explanation, the NIEDF modeling developed in 11 and 25 was applied for several values of hydrogen coverage on carbon surface (set as input to SRIM software) and compared to the experimental data for various pulsed-bias frequencies. As one can see from Figure 2 (a), the hydrogen coverage on HOPG surface is estimated to decrease from 40% to 30% when increasing the frequency from 5 kHz to 90 kHz. The value of H coverage for DC-biased HOPG is estimated to be 30%[25]. Therefore, we can conclude that the increase of the measured signal with the decrease of the pulse-bias frequency is due to increase of the hydrogen coverage on the surface. A bigger hydrogen surface coverage increases the number of NI created by sputtering, and may also change the electronic properties of the top surface material thereby affecting the surface ionization probability[24] (the probability for a H particle to capture and keep an electron while leaving the surface).

The increased surface coverage can be understood as follows. Surface coverage is increased by atoms and ions impinging on the surface and decreased by the self-sputtering (sputtering of hydrogen atoms by hydrogen ions) and by physico-chemical etching. The self-sputtering is not continuous in this case (and most probably the physico-chemical etching is also time modulated), so there is time for atomic hydrogen to fill the surface up to the saturation if the bias **OFF** period is sufficiently long. That is probably why the NI signal does not decrease much until 1 kHz. On the other hand, as we approach to the continuous bias situation while increasing the bias frequency, we leave less time for H atoms to come back to the surface. So, the coverage gets smaller until it reaches the continuous DC-bias value.

Let us note that the pulsed bias method does not represent a way to increase the production of NI since we measure the NI yield only during the ON pulse duration. The total production in pulsed mode is the product of the production during the **ON** pulse duration by the duty cycle (assuming no production during the **OFF** duration), which is well below the production in continuous mode.

## 5. Results for an insulating material: MCD

No NI signal could be measured with MCD sample using a continuous DC bias since the sample loses its surface conductivity and becomes insulating when exposed to hydrogen plasma at room temperature. Therefore, the pulsed DC bias method has been applied using a 100 μs DC pulse ($T_{pulse}$ = *100 μs*) with a repetition frequency of 1 kHz (duty cycle = 10 %). The acquisition time has been set to 10 μs ($T_{acq}$ = *10 μs*) and the delay was 13 μs ($T_{delay}$ = *13 μs*). As it can be seen in Figure 3 (a) NIEDF measurement can be recovered using the pulse DC bias method. The onset of the NIEDF is at 0 on the energy scale indicating that for this measurement the surface bias is indeed equal to –130 V. Let us note that a bias $V_a$ equal to -140 V has been applied to get a surface bias of about -130 V.

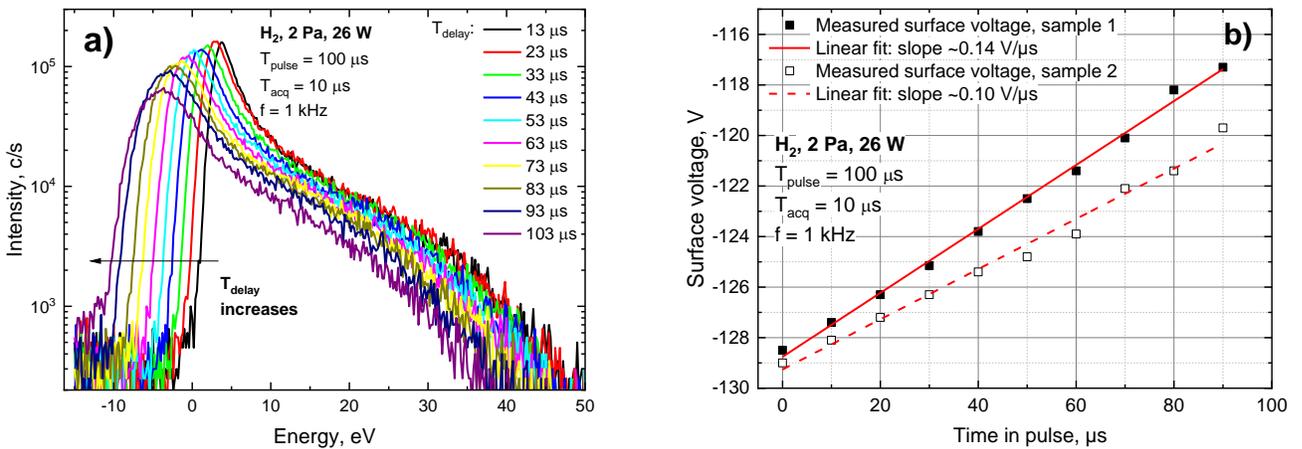

Figure 3 : 2.0 Pa of $H_2$, $P_{RF}$ = 26 W, $V_a$ = -140 V (a) NIEDF measured at different moments (defined by $T_{delay}$) of the bias pulse. The pulse duration is $T_{pulse}$ = 100 μs, the acquisition duration is $T_{acq}$ = 10 μs. $T_{delay}$ is varied with a step of 10 μs from the beginning until the end of the pulse. Bias frequency is 1 kHz, ceramic clamp, sample 1 (b) surface voltage $V_s$ of MCD material at different moments of the 100 μs bias pulse for two different sample thicknesses (sample 1 is 17.5 μm thick whereas sample 2 is 12.5 μm thick). Surface voltage are deduced from the peak shift observed in the NIEDFs (shown in Figure (a) for sample 1). NIEDFs have been measured with a ceramic clamp for sample 1 and a molybdenum clamp for sample 2.

To study the charge accumulation on the surface during the bias pulse, NIEDF acquisitions have been performed at different delay times ($T_{delay}$ from 13 to 103 μs) and are presented in Figure 3 (a). All NIEDF acquisitions have been processed assuming a surface bias of -130 V (according to equation 5). NIEDF peak positions as well as NIEDF onsets are shifting towards negative values indicating that the surface bias is increasing and is no more equal to -130 V. This shift is equal to the difference between the actual surface bias and the bias initially present on the surface ($V_s \sim -130\ V$). The NIEDF onset has therefore been used to get the actual surface bias $V_s$ present on the surface at different times in the pulse (different $T_{delay}$). In such a way, the charging of the sample was followed in real time as demonstrated in Figure 3 (b). One can observe that charging of MCD is happening as expected in a linear manner with a rate of ~$1V/10\ \mu s$. In Figure 3 (b) the variation of the surface bias is plotted as a function of time for a ceramic clamp and a sample of 17.5 μm thickness, and for a molybdenum clamp and a sample of thickness 12.5 μm. In the case of the metallic clamp, the voltage on the holder surrounding the sample remains at -140 V while the voltage on the sample varies from -130 V to -120 V. With the ceramic clamp, the voltage on the holder surrounding the sample is always at floating potential (few volts above 0) during the measurements. However, no noticeable effect of the holder material was observed. Furthermore, the slopes obtained are in the ratio of sample thicknesses which reinforces the simple picture of sample charging presented in the third part of this paper.

The charging of the sample during the pulse increases the value of bias voltage $V_s$ on the surface. Therefore, $T_{acq}$ must be sufficiently short to get undistorted NIEDF shape which can be compared to NIEDF obtained with conductive samples biased at a well-defined voltage. The acquisition time of $T_{acq} = 10\ \mu s$ seems reasonable to provide a quasi-constant surface bias $V_s$ during measurement ($V_s$ increases by only *1 V* out of *130 V* during the measurement).

To get information about accumulated charges on the surface during the pulse and their removal between pulses (the OFF time), study of the bias duration influence on the NIEDF measurements was performed and is presented in Figure 4. The duration of acquisition was kept constant $T_{acq} = 8\ \mu s$ with a delay $T_{delay} = 13\ \mu s$. Therefore, only the first ~10 μs of the pulse are probed. From the NIEDF onset shift of Figure 4 (a), the surface bias is deduced and plotted as a function of $T_{pulse}$ in Figure 4 (b), for both clamps. No noticeable effect of the clamp material is observed showing that the sheath in front of the holder surrounding the sample is not affecting the measurements. The variations of the NIEDF intensity observed in Figure 4 (a) with the bias duration will be explained later.

The surface voltage at the beginning of the pulse is almost constant with $T_{pulse}$ up to the bias duration of *100 μs* (duty cycle of *10%*) or for pulse-off duration $T_{off}$ >*900 μs*. After that, the surface voltage varies fast with $T_{pulse}$, indicating that the charge on the surface has not been eliminated between two pulses. Indeed, if the OFF duration is not long enough and/or if the ON duration is too long, corresponding in both cases to a high duty cycle, there is a residual charge on the surface at the beginning of next pulse and the surface bias is no more equal to the applied bias plus the undisturbed floating potential (equation 4).

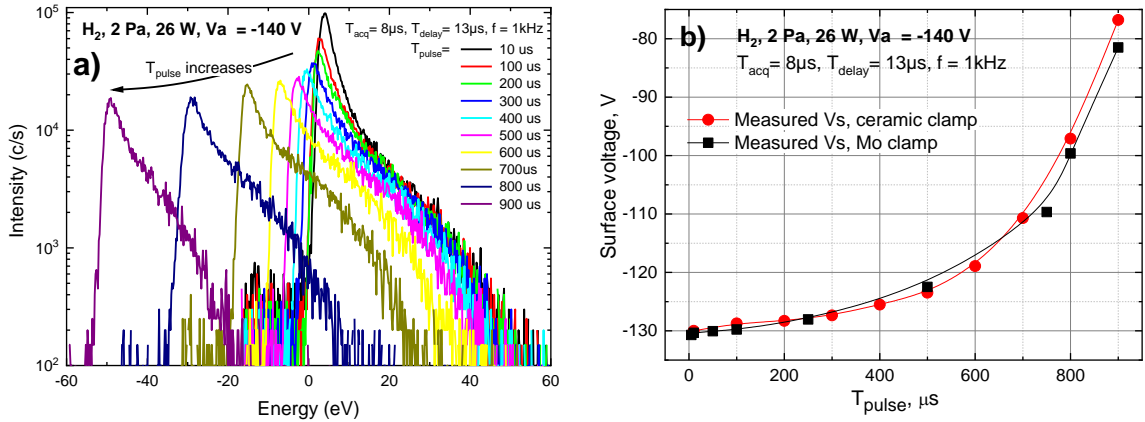

Figure 4: Influence of pulse duration ($T_{pulse}$) on surface voltage $V_s$ for MCD surface at RT in $H_2$ plasma, 26 W, 2 Pa, $V_a$ = –140 V, f = 1 kHz, Tacq = 8 μs and Tdelay = 13 μs (a) NIEDF measured at beginning of pulse for pulse duration $T_{pulse}$ from 10 μs to 900 μs, ceramic clamp (b) Vs at beginning of pulse as a function of $T_{pulse}$ with a ceramic clamp (circle symbols, deduced from a) and a molybdenum clamp (square symbols)

To understand the behavior of $V_s$ shift with bias pulse duration, one can refer to the scheme shown in Figure 1 (a). If the pulse is short enough ($T_{pulse}$ < 100 μs) the accumulated charge during the bias pulse leads to a positive surface bias of only few volts when the applied bias is switched off. This value is close to the floating potential (~ 6 V), therefore, no significant perturbation of the plasma is induced at the end of the pulse. If the pulse is long the accumulated charge leads to high positive bias on the surface when the applied bias is switched off, on the order of 20 volts for $T_{pulse}$ = 200 μs and 90 V $T_{pulse}$ = 900 μs. Therefore, after the pulse, the surface appears to have a large positive bias $V_s > V_p$. Electrons are attracted towards the sample resulting in a high loss of electrons and a consequent increase of the plasma potential. This results in a quick initial discharge of the surface (electrons are attracted) and then (for $V_s \leq V_p$) the discharging happens on a longer time scale, as illustrated in Figure 1 (a). To support this explanation, positive-ion energy distribution functions (PIEDF) have been measured just before the pulse and just after the pulse for two bias durations 50 and 500 μs. $H_3^+$ is the dominant positive ion, its transit time to the mass spectrometer detector is 19 μs. Therefore, $T_{delay} = 0$ μs in the figures corresponds to a PIEDF measurement ~ 20 μs before the bias pulse is applied.

Positive ion energy distribution for the case of $T_{pulse} = 50$ μs with $T_{acq} = 5$ μs just before the pulse is shown in Figure 5 (a). As expected, it coincides well with the curve for no bias, meaning that no effect from the previous pulse is present in the plasma. The other curve represents the situation for $T_{delay} = 70$ μs and demonstrates that immediately after the bias pulse, the PIEDF is identical to the one in the absence of bias. It shows no disturbance of the plasma potential by the accumulated charge on the surface.

The situation for the long bias pulse is shown in Figure 5 (b). In this case, the charge accumulated on the insulating surface leads to a voltage on the order of 50 V, so $T_{off}$ = *500 µs* appears to be too short to discharge it completely. Energetic part of PIEDF immediately before the pulse ($T_{delay}$ = *0 µs*) is not strictly identical to the one with no bias. The measurements after the pulse ($T_{delay}$ > *520 µs*) show that some high-energy $H_3^+$ are present in the PIEDF, demonstrating that the distribution of plasma potential has been disturbed. Most probably, the plasma potential close to the sample is increased in order to shield the positive surface charge and limit the loss of electrons. This high energy ion component of the PIEDF slowly decreases during the **OFF** time, demonstrating a slow decrease of the surface charge accompanied by a slow decrease of the plasma potential.

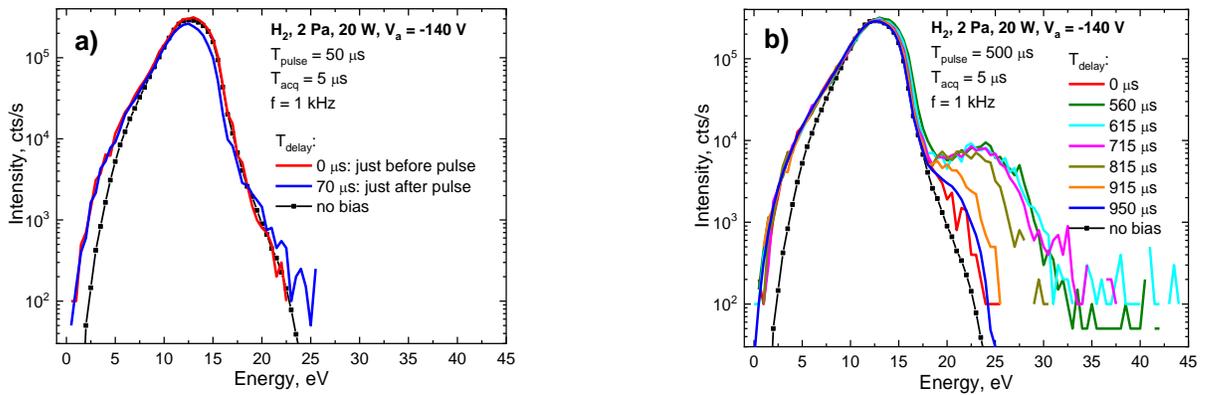

Figure 5: PIEDF of $H_3^+$ with (a) $T_{pulse}$ = 50 µs (b) Tpulse = 500 µs and $T_{acq}$ = 5 µs for various values of $T_{delay}$ compared with the $H_3^+$ PIEDF in the absence of applied bias (symbols). Plasma parameters: 2.0 Pa of $H_2$, $P_{RF}$ = 20 W, $V_a$ = -140 V, molybdenum clamp.

This study demonstrates the complex surface charge dynamics during the **OFF** duration, and possible plasma potential disturbances due to the accumulated charge on the surface. The same effect can also be observed by varying bias frequency and keeping pulse duration constant. The measurements shown in Figure 6 (a) were made with an applied bias $V_a$ =*−140 V* and for $T_{pulse}$ = *10 µs* at different frequencies from 0.25 to *90 kHz*, which gives duty cycle from *0.25%* to *90%* and OFF duration from 3990 µs to 1 µs correspondingly. As observed in Figure 6 (a) the NIEDF is shifted and its intensity is decreased. For high frequencies (high duty cycle) the MCD sample cannot discharge completely during $T_{off}$. It means that for the bias duration used in the experiment ($T_{pulse}$ = *10 µs*), the frequency should not be higher than *5–10 kHz* (see Figure 6 (b) square symbols) to get a surface bias of *–130 V* with an applied bias of *–140 V*. For higher frequency, the applied bias must be increased to get the appropriate surface bias.

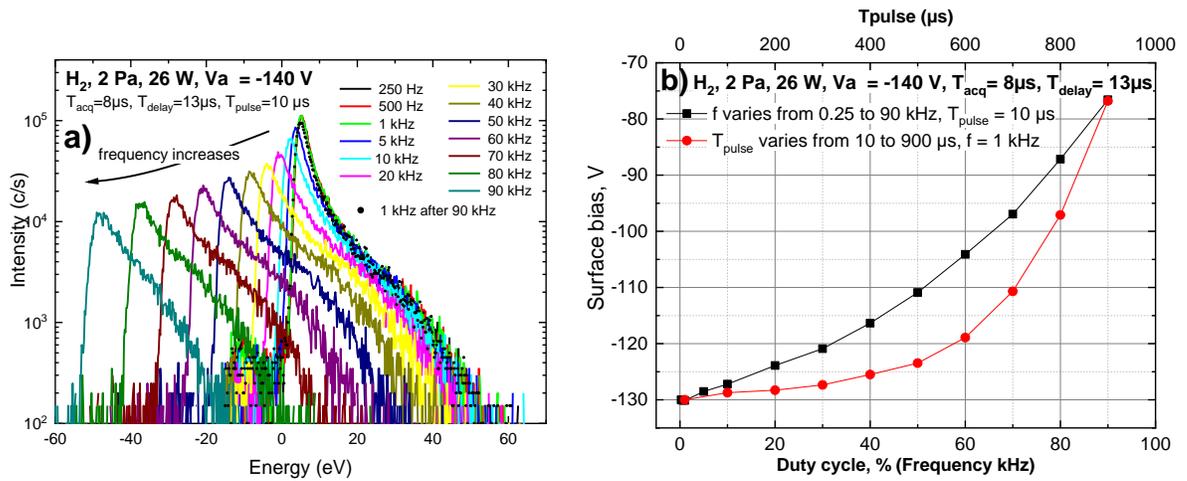

Figure 6 $H_2$, 26 W, 2 Pa, $V_a$ = –140 V, $T_{acq}$ = 8 μs and $T_{delay}$ = 13 μs, ceramic clamp (a) NIEDF measured during a 10 μs pulse ($T_{pulse}$ = 10 μs) at different frequencies from f = 0.25 kHz to 90 kHz with $T_{acq}$ = 8 μs and $T_{delay}$ = 13 μs (b) $V_s$ at beginning of pulse as a function of duty cycle when varying frequency (square symbols deduced from Figure 6 (a) at constant pulse duration ($T_{pulse}$ = 10 μs), and when varying pulse duration (round symbols deduced from Figure 4 (a) from 10 to 900 μs at constant frequency (f = 1 kHz)

In Figure 6 (b) the surface bias $V_s$ is plotted versus the duty cycle for both experiments, when varying pulse duration at constant frequency ($V_s$ deduced from Figure 4 (a), top horizontal axis), and when varying frequency at constant pulse duration ($V_s$ deduced from Figure 6 (a), bottom horizontal axis). First, as expected the duty cycle is a key parameter. The higher the duty cycle, the greater the charge on the surface, the lower the surface bias (in absolute value) during the pulse for a given applied voltage. Second, the discharge rate of the surface is clearly slower than the charging rate, otherwise up to a duty cycle of 50 % the surface bias would be -130 V. Third, the discharge rate is non-linear with the charge of the surface. It is higher for a higher charge, otherwise no negative stationary surface bias would be reached for high duty cycle and the sample voltage would reach the floating potential. The non-linearity comes from the electron current $I_e$ discharging the sample which is an increasing function of the sample bias $V_s$ ($V_s$ at the beginning of the OFF period is proportional to the accumulated charge on the sample). The more the accumulated charge on the surface is, the higher is the initial electron current discharging the sample. Due to this non-linearity, the two type of experiments do not give the same surface bias for the same duty cycle. Indeed, the two curves on the plot cover very different situations. A duty cycle of 50% corresponds to 500 μs of bias ON, 500 μs of bias OFF in one situation, and to 10 μs ON and 10 μs OFF in the other situation (frequency variation), leading to two different surface biases at the beginning of the pulse ∼-124 V ($T_{pulse}$ variation) and ∼-111 V (frequency variation). As the charging rate is expected to be the same and equal to $I_i/C$ in both experiments, it appears that the discharge rate is faster in the situation with 500 µs ON, 500 µs OFF. Most probably, the high charge of the surface in this situation ($\Delta V \sim 50$ V at the end of the pulse, Figure 3) is provoking a huge flux of electrons in the first instants of the OFF period that might lead to a fast initial discharge of the surface. At low duty cycle there is no difference between both experiments since the OFF period is long enough in both cases to completely discharge the surface (990 µs OFF in the case of $T_{pulse}$ variation, 4 ms in the case of frequency variation). However, it is not clear why both experiments merge at high duty cycle. Future works will focus on sample current measurements during ON and OFF phase to better understand charging and discharging of the sample surface.

One can note in Figure 6 (a) that NIEDF shapes are modified as the pulse bias frequency grows. They are quite usual and similar to the ones obtained on conductive samples (see Figure 2 (a) or reference 9) at low frequency and broaden and lose the marked low energy peak at higher frequency. At the same time the NI signal (area below NIEDF) noticeably decreases. This cannot be interpreted easily because, as the surface voltage is changing, the mass spectrometer tuning is not optimized anymore for the correct NI energy. To prevent the signal loss resulting from unsuitable tuning, the applied bias $V_a$ has been increased in order to have *$V_s$ = -130 V* on the sample surface whatever the pulse frequency. In such a way, the situation was comparable to the use of a conducting sample as the surface bias remained at desired value. As can be seen from Figure 7 (a), the tails of measured NIEDF superimpose up to *f = 90 kHz* and only the distribution maximum value decreases. This may indicate the decrease of the sputtering contribution to the NI production resulting from the hydrogen surface coverage decrease. However, the suppression of the sputtering mechanism could account at maximum for a decrease of the signal by a factor two[11]. Therefore the decrease of the hydrogen surface coverage may also lead to a change of top surface electronic properties and thus to a change of the ionization probability[24]. A modification of incoming positive ion flux and energy for short $T_{off}$ (high frequency) cannot be excluded as well but has not been investigated here. It will be the subject of future works. Let us note that NIEDF modelling cannot be used here since the model assumes a planar sheath in front of the sample to calculate negative-ion transport to the mass spectrometer[22]. In the present situation, because of the ceramic clamp surrounding the sample, the sheath cannot be planar. In the case of the molybdenum clamp the sheath cannot be planar either since the applied bias appears on the clamp while only the surface bias appears on the sample. The same experiment done with the molybdenum clamp leads to basically the same results, with similar values of the applied biases at each frequency, and a more pronounced decrease of the signal probably due to higher distortion of the sheath in front of the sample. Finally, one can note that adjusting the applied bias to get a constant surface bias at any frequency has allowed solving the issue of inappropriate mass spectrometer tuning and has somewhat increased the measured signal (Figure 7 (b)).

To study negative-ion production on MCD sample under the present experimental conditions, a pulse duration of 10 µs is appropriate since it leads to a small change of surface bias during the mass spectrometer acquisition (1 V out of 130 V). The pulse frequency has to be chosen as high as possible to minimize measurement time and as low as possible to increase negative-ion signal and get undisturbed NIEDF (Figure 7 (a)). Using 1 kHz seems to be a good compromise. These parameters will be used in the following.

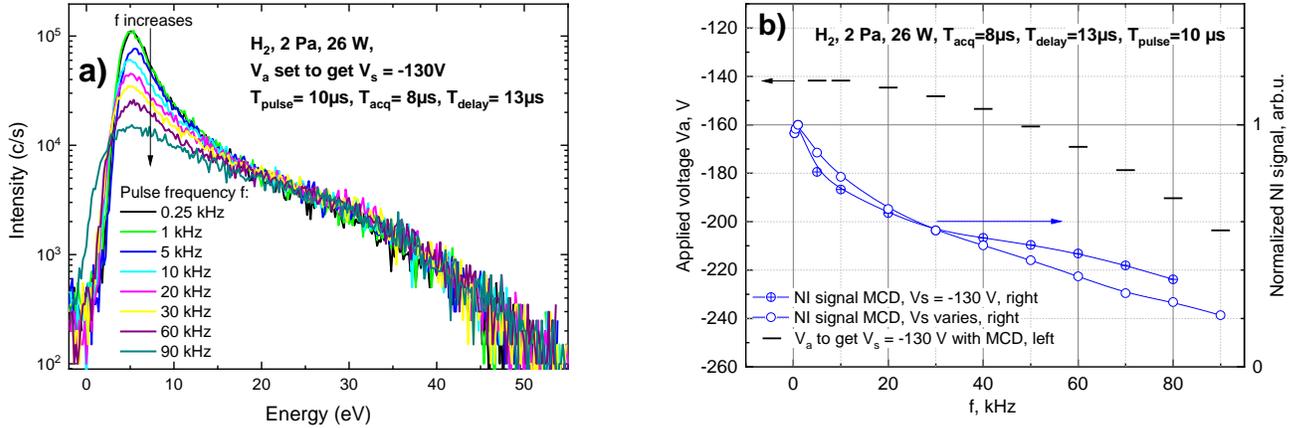

Figure 7: $H_2$, 26 W, 2 Pa, Tpulse = 10 µs, Tacq = 8 µs and Tdelay = 13 µs, ceramic clamp (a) NIEDF versus pulse frequency from f = 0.25 kHz to 90 kHz with applied voltage set to maintain surface voltage equal to -130V (b) Normalized NI total signal versus frequency for MCD with Vs = -130 V (cross-circle symbols, deduced from Figure 7 (a), right scale), MCD with Vs varying (open circle symbols, deduced from Figure 6 (a), right scale), and value of the applied voltage set to get surface voltage of -130V (line symbols, left scale)

## 6. Results for heated materials

Top layers of samples exposed to plasma are modified by the ion and atomic bombardment. At low temperature the top layer formed on MCD sample is probably insulating and does not permit to bias the sample. As observed previously[17], at higher temperature, above 300 °C, this top layer becomes conductive and the sample can be biased through the clamp contact. The pulsed bias method allows studying negative-ion surface production on MCD sample on the whole temperature range without any restriction and allows comparing directly pulsed and DC bias at high temperature. Previous works have shown that the total NI count on different kind of diamond samples is increasing with temperature reaching a maximum around 400-500 °C and decreasing above. This has been attributed to the enhanced etching of sp2 carbon phases at higher temperature allowing to obtain a top layer which is closer to the pristine diamond sample. Recovering the diamond electronic properties seems therefore to be favorable for negative-ion surface production. It has also been noticed that the hydrogenation of the top layer is increasing between room temperature and 400-500 °C and decreasing above. The decrease of the signal has been attributed to the decrease of the hydrogenation, leading to a decrease of the negative-ion creation by sputtering, and also probably to a change of electronic properties which may become less favorable to surface ionization.

NI total signal versus surface temperature for HOPG, MCBDD and MCD materials are presented in Figure 8. For MCD and MCBDD materials two samples from the same batches have been used in two different experimental campaigns. The signal from the two campaigns are sometimes slightly different which might come from different alignment of the sample in front of the mass spectrometer or a problem of temperature calibration. As the temperature is measured on the sample holder and not on the sample itself, the surface temperature might be different from the indicated one due to variations of thermal contact. One can note that MCD samples are giving lower signal than MCBDD ones in several situations, in particular in DC mode. We do not think that this has something to do with the doping as MCD material has already demonstrated identical negative-ion yields to MCBDD one[9,17]. The most remarkable difference between the MCD layer used in the present study and MCD material used in previous works is the size of the diamond grains which are ~10 µm in the horizontal plane here and were around 2.5 µm previously. However, the exact reason for the lower yield is outside the scope of this paper. It will be the subject of future works.

From Figure 8 one can observe that the pulsed bias mode leads to a higher negative-ion signal for all materials, with a huge increase compared to DC bias mode, almost one order of magnitude, for diamond materials. For all materials the hydrogenation of the top layer, deduced from the modeling, is higher in pulsed mode than in DC bias mode. However, this is not enough to explain the large signal increase observed for diamond materials. For instance, at room temperature signal on MCBDD material is lower than signal on HOPG while it becomes higher by a factor 4 in pulsed mode. This large increase of the signal is attributed to the low number of defects created on diamond when exposed to pulsed DC bias at low duty cycle (1% here). The diamond surface subjected to pulsed bias is close to the pristine diamond surface which electronic properties are probably favorable to surface ionization of outgoing hydrogen particles. In pulsed mode one acquisition lasts around 20 minutes. It represents about 12 seconds of positive-ion bombardment and therefore measurements in pulsed mode can be considered as done with a "poorly defective" material. Furthermore, atomic bombardment during the OFF phases (without bias) might help removing defects created during the ON phases. The situation is close to the one obtained at the first instants of plasma exposure where the surface is not too much damaged and the signal is found higher than at steady state[9]. This interpretation is supported by Figure 8 (b) presenting MCD total counts versus temperature in pulsed and DC mode, and showing the timing of the experiment. The measurements are first done in pulsed mode up to 600°C and then back to 400°C. No hysteresis is observed. As the top MCD surface is conductive at 400°C the bias was then switched to DC mode. The graph shows that the signal is decreasing with time to reach a steady state after around 15-20 minutes with a much lower signal. This decrease is attributed to defect creation by the positive ion bombardment. Considering the low positive ion flux in this experiment, a fluence of $3 \cdot 10^{15}$ ions/cm$^2$ (roughly three mono layers of material) is reached after about 1 minute, and this slow evolution over 20 minutes is not surprising. Then the pulsed bias is turned ON again and a memory effect is observed. The signal is increasing, probably due to the higher material hydrogenation, but not as much as it was before DC bias irradiation. This is attributed to the defects created during the DC bias exposure.

Both in pulsed and DC modes the negative-ion signal is increasing with temperature for diamond materials. This is attributed to the enhanced sp2 phases etching at higher temperature. The decrease above 400-500°C is explained as in DC bias mode by the decrease of the material hydrogenation[9,23,24]. Concerning HOPG the results are very similar in pulsed and DC bias modes, except for the increase of the signal at room temperature in pulsed mode due to the higher material hydrogenation. At higher temperature, as shown previously, hydrogen is desorbing, and the graphite material is recovered, leading to a strong decrease of the signal[9,23,24].

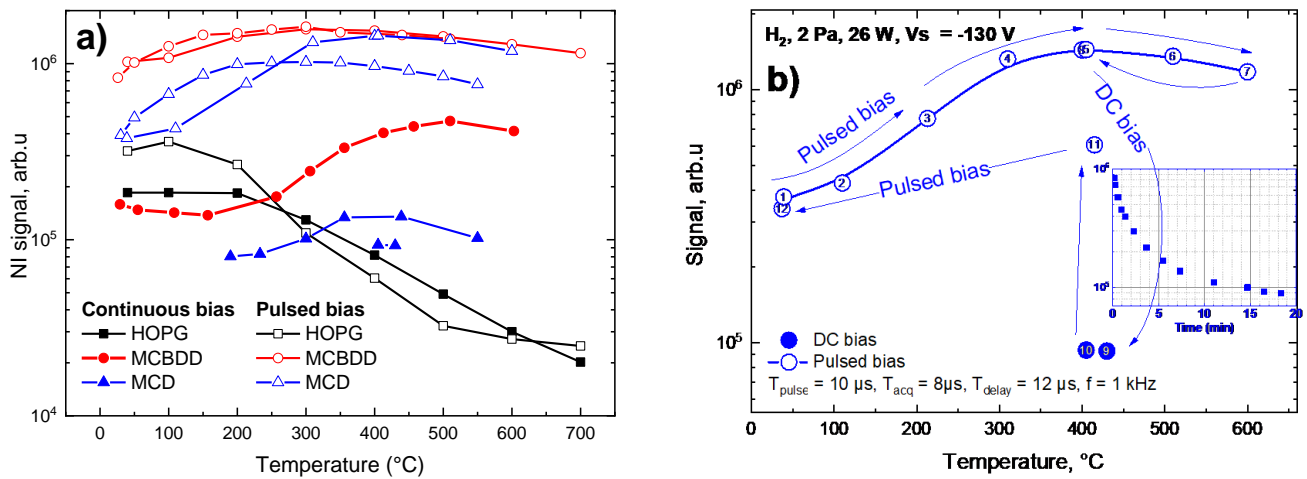

Figure 8 : H$_2$, 26 W, 2 Pa, Vs = -130V, pulsed bias conditions: Tpulse = 10 µs, Tacq = 8 µs and Tdelay = 13 µs, 1 kHz, molybdenum clamp. Negative ion total signal versus surface temperature in DC (full symbols) and pulsed mode (open symbols) (a) for three different materials: HOPG, MCBDD and MCD (b) for MCD material with the timing of the experiment. The numeric labels correspond to the chronological order of the experiment. The insert shows NI signal versus time between experiment #8 and #9.

# 7. Conclusion

In previous works we have developed experimental methods to study NI surface production in plasmas. A sample is immersed in hydrogen plasma and is negatively biased. Upon positive ion bombardment some negative-ions are formed on the surface and are accelerated towards the plasma. Thanks to the energy gained in the sheath they can be self-extracted on the other side of the plasma towards an energy and mass analyzer. Up to now this method was limited to conductive materials due to the use of a DC bias. In the present paper, the method of pulsed bias was developed to enable the study of NI production on surfaces of insulating materials. The idea is to apply a brief DC bias pulse on the surface and perform a synchronized negative-ion acquisition. Between pulses the surface is kept unbiased to ensure discharging of the charges accumulated during the pulse. The present technique has enabled to study NI surface production on MicroCrystalline Diamond (MCD) for a temperature range starting from RT to 700°C. The pulsed-bias tests were first performed on Highly Oriented Pyrolitic Graphite (HOPG), a conductive material, to demonstrate the feasibility of the method. By changing the pulsed-bias frequency (and the duty cycle) it was possible to obtain HOPG material with different hydrogen surface coverage and hence a different surface state (with the NI yield increase up to *30-50%*). After proving the feasibility of the pulsed bias approach on HOPG, the optimization of the experimental parameters was performed on MCD by taking into account the charging effects. It has been shown that the MCD sample behaves in plasma as a capacitance charged by a constant current (the positive ion saturation current) during the pulse. A low duty cycle is required (<10 %) to ensure between bias pulses a complete removal of charges accumulated during the pulse. A pulse duration of 10 µs with a repetition frequency of 1 kHz allows for measurements on MCD material under the present low positive-ion flux experimental conditions.

The total negative-ion signal strongly increases between DC and pulsed bias modes, for the three materials tested: HOPG, MicroCrystalline Boron Doped Diamond and MCD. Factors from 1.5 (HOPG) to around 10 have been observed and the highest negative-ion yield ever measured under the present experimental conditions have been obtained in pulsed mode at high temperature. The results bring us to the conclusion that in pulsed bias case the diamond surface is less degraded and more hydrogenated, which is favorable for NI surface production. This situation is similar to biasing a fresh diamond sample with constant DC bias for a very short exposure time. Most probably, under ion bombardment, diamond samples lose their attractive electronic properties. With a very short exposure time, or using the pulsed bias technique, it is possible to maintain electronic properties close to those of pristine diamond and obtain higher NI yields. Recent measurements[29] show that relative negative-ion yield measured by the mass spectrometer could be placed on an absolute scale by using a Magnetic Retarding Field Energy Analyzer (MRFEA)[30]. This will be the subject of future work. However, by comparison with our previous measurements we can already state that NI yields on diamond measured in the present paper are probably two orders of magnitude higher than on metallic surfaces such as molybdenum[31,32].

Pulsing the bias at low duty cycle is a solution only for fundamental studies and could not be used in a real NI source due to the low time average negative-ion production. However, the present study shows that electronic properties of diamond are favorable for NI production, and diamond placed in appropriate experimental conditions could be potentially interesting for NI surface production as also shown in beam experiments[22]. Considering the overall result of the pulsed bias measurements, one can conclude that to optimize the NI yield on diamond, one has to work with a less degraded surface. This can be obtained by rising the surface temperature to 400°C–500°C which allows restoring intrinsic properties of diamond. The less degraded surface state can also be obtained by reducing positive ion energy.

To conclude, this works shows that non-degraded diamond surface, or more generally insulator materials with electronic properties similar to those of diamonds (band gap, possibly negative electron affinity...), are promising candidates to enhance surface NI production in NI sources. Pulsed DC bias is an efficient experimental method developed for fundamental studies but could be replaced by tailored waveform bias in practical applications[15,33,34].

## Acknowledgments

This work was carried out within the framework of the French Research Federation for Fusion Studies (FR-FCM) and the EUROfusion Consortium and has received funding from the Euratom research and training programme 2014-2018 under grant agreement No 633053. The views and opinions expressed herein do not necessarily reflect those of the European Commission. Financial support was received from the French Research Agency within the framework of the project H INDEX TRIPLED 13-BS09-0017. PACA county is gratefully acknowledged for its financial support through project 'PACAGING 2012_10357'. CGI (Commissariat à l'Investissement d'Avenir) is gratefully acknowledged for its financial support through Labex SEAM (Science and Engineering for Advanced Materials and devices) ANR 11 LABX 086, IDEX 05 02